\documentclass[twocolumn,showpacs,preprintnumbers,amsmath,amssymb]{revtex4}
\usepackage{graphicx}
\usepackage{dcolumn}
\usepackage{bm}

\begin{document}

\title{Carbon nanotubes on partially depassivated \emph{n}-doped Si(100)-(2$\times$1):H substrates}

\author{Salvador Barraza-Lopez$^1$}
\email{sbl3@mail.gatech.edu}
\affiliation{
1. School of Physics. Georgia Institute of Technology. Atlanta GA, 30332.\\
2. Department of Electrical and Computer Engineering and Beckman Institute for Advanced Science and Technology. University of Illinois. Urbana IL, 61801.}
\author{Peter M. Albrecht$^2$}
\email{albrecht@engineering.uiuc.edu}
\affiliation{
1. School of Physics. Georgia Institute of Technology. Atlanta GA, 30332.\\
2. Department of Electrical and Computer Engineering and Beckman Institute for Advanced Science and Technology. University of Illinois. Urbana IL, 61801.}
\author{Joseph W. Lyding$^2$}
\affiliation{
1. School of Physics. Georgia Institute of Technology. Atlanta GA, 30332.\\
2. Department of Electrical and Computer Engineering and Beckman Institute for Advanced Science and Technology. University of Illinois. Urbana IL, 61801.}
\date{\today}

\begin{abstract}
 We present a study on the mechanical configuration and the electronic properties of semiconducting carbon nanotubes supported by partially depassivated silicon substrates, as
inferred from topographic and spectroscopic data acquired with a room-temperature ultrahigh vacuum scanning tunneling microscope
and density-functional theory calculations. A mechanical distortion and doping for semiconducting carbon nanotubes on Si(100)-(2$\times$1):H with
hydrogen-depassivated stripes up to 100 \AA{} wide are ascertained from both experiment and theory. The results presented here point towards novel 
and local functionalities of nanotube-semiconductor interfaces. 
\end{abstract}

\maketitle

The modifications of the intrinsic electronic and thermal properties of single-walled carbon nanotubes (SWNTs) due to their interaction with the semiconducting surface by which they are 
supported have been the focal point of a sizeable number of experimental\cite{a1,a2,a4,a6,a7,a8,aplrecent,peternanotechnology,a18} 
and theoretical\cite{a8,a9,a13,a17,GalliPRL} studies in recent 
years due to the technological interest in hybrid SWNT-semiconductor devices. Dry contact transfer 
(DCT)\cite{a2} allows for the \emph{in situ} deposition of SWNTs from solid sources onto 
technologically relevant surfaces, such as Si(100)\cite{a2,a6,a8,peternanotechnology,a18} and 
the (110) surfaces of GaAs and InAs\cite{a4}, forming 
 an atomically pristine interface.
 Comprehensive studies of semiconducting SWNTs (s-SWNTs) on Si(100) and Si(100)-(2$\times$1):H at the density-functional theory (DFT) level
have been reported\cite{a17,a8}. The adsorption properties of s-SWNTs and metallic SWNTs (m-SWNTs) of 
similar diameter on Si(100) are remarkably different\cite{a17}. This holds for $\sim$10 \AA-diameter SWNTs with 
varying chiralities. Unlike m-SWNTs on Si(100), no covalent bonds are formed at the 
interface between Si(100) and s-SWNTs. (In recent studies of a (10,0) semiconducting SWNT on Si(100), covalent bonds between the s-SWNT and Si 
surface atoms were artificially induced\cite{GalliPRL}
.) 
 Si-C covalent bonding for SWNTs grown on Si was asserted only for a subset of the population 
based upon micro-Raman imaging\cite{aplrecent}, suggesting that both the electronic character of the 
SWNT and the local termination of the Si surface markedly influence the degree of nanotube-substrate interaction. When 
SWNTs are placed on the inert monohydride Si(100)-(2$\times$1):H surface\cite{a2}, 
 the respective electronic characteristics of the nanotube and the surface are preserved and charge transfer is largely 
suppressed\cite{a8, a13}. In contrast, a shift of the Fermi level away from the nanotube midgap position has been experimentally observed for SWNTs
 supported by metallic surfaces (Au(111)\cite{DekkerLieber} and Ag(100)\cite{Kawai}), unpassivated III-V compound semiconductors (GaAs(110) 
and InAs(110)\cite{a4}) and an ultrathin insulating film (NaCl(100)/Ag(100)\cite{Kawai}).
 Faithful to experimental 
conditions, our computational studies included dopants within the Si(100) slab. 
 Ultrahigh vacuum scanning tunneling microscope (UHV-STM) based nanolithography on
hydrogen-passivated Si(100)
 enables the definition of patterns of reactive depassivated Si\cite{Lyding1,Lyding2} with possible consequences for
the adhesion and electronic properties of the adsorbed SWNTs\cite{a6,peternanotechnology,a18}.
 In this Letter, we report on the properties of isolated s-SWNTs interfaced with nanoscale regions of selectively depassivated Si as determined
 from room-temperature UHV-STM measurements and DFT calculations.
\begin{figure*}
\includegraphics[width=.8\textwidth]{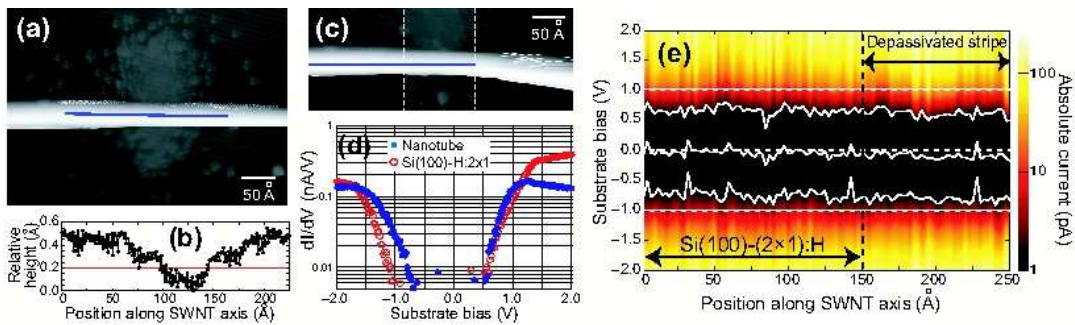}
\caption{\label{fig:fig1}(color online) (a) UHV-STM image ($-2$ V, 100 pA) of a SWNT interfaced with a partially depassivated 
Si(100)-(2$\times$1):H substrate. Darker (brighter) areas on the substrate indicate hydrogen passivation (depassivation), with the SWNT seen as a horizontal white 
feature. (b) STM constant-current height profile along the SWNT (blue line in (a)) revealing a sub-0.5 \AA{} depression. Spectral data 
were subsequently acquired along the blue line in (c). The residual mechanical instability of the nanotube on H-Si to the right of the second vertical line 
manifests as a broadening of the apparent width of the SWNT. 
(d) Semilogarithmic differential conductance profile (acquired on the SWNT over the Si(100)-(2$\times$1):H area) indicates 
the SWNT is semiconducting; see Ref. \onlinecite{a8}. The $n-$type alignment 
of the substrate d$I$/d$V$ is also evident. (e) Absolute tunneling current plot along the axis of the SWNT (blue line in Fig.~1(c). Spatial resolution: 2.5 \AA.). To 
the right of the vertical dashed white line the 
tube sits on the depassivated stripe. The white traces in the profile signal a current of 1 pA, (close to the edge of the valence and 
conduction bands). Horizontal dotted lines serve as guide to the eye. The midgap energy (also shown) depicts a 
slight $n$-doping of the SWNT in the section on top of the depassivated substrate.}
\end{figure*}
\begin{figure}
\includegraphics[width=.45\textwidth]{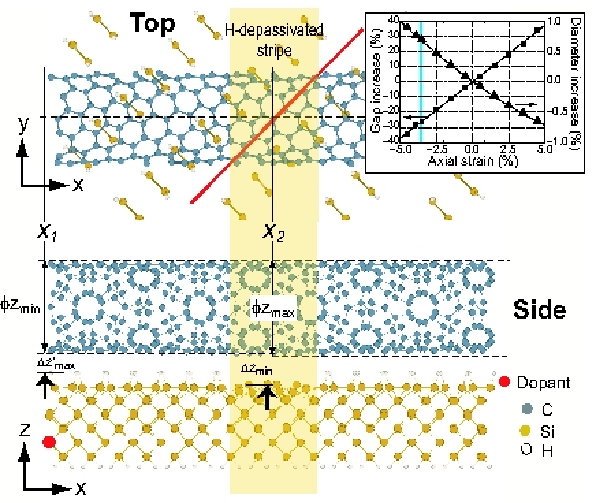}
\caption{\label{fig:Fig2} (color online) Top: Ball-and-stick model of the (8,4) SWNT on a partially depassivated (10 \AA{} wide) $n$-doped Si(100)-(2$\times$1):H substrate. The red
line indicates the direction of the Si dimmer rows, and it forms a 45$^{o}$ angle with respect to the SWNT axis. Only the lower half of the SWNT is
shown for clarity. Side: The side view of the system, showing in red the location of the phosporous ($n-$type) dopant. 
Horizontal dashed lines serve to
indicate distortion in the SWNT. Maximum distortion in SWNT occurs at
$x_1$ and $x_2$. Inset: The (8,4) SWNT was contracted by 3.8\% (blue vertical line) in order to be placed on top of the Si substrate. This
implied a 30\% reduction of its gap and a 0.75\% increase in diameter to compensate longitudinal compression.}
\end{figure}

Fig.~\ref{fig:fig1} summarizes the experimental observations; reproducible results were obtained for several unique s-SWNTs on partially depassivated Si(100)-(2$\times$1):H.
 Degenerately $n$-type doped Si(100) substrates (As, $10^{19}$ cm$^{-3}$)
were employed, and subjected to 
UHV H-passivation\cite{a2}. Isolated HiPco SWNTs\cite{FN1} 
were deposited by DCT. The STM was operated at room temperature in constant-current mode 
with the bias voltage ($V$) applied to the substrate and the electrochemically etched W tip grounded through a 
current ($I$) preamplifier. Partial surface depassivation, as seen 
in the filled-states topograph of Fig.~1(a), was achieved with the methods described in Refs.~\onlinecite{
peternanotechnology} and \onlinecite{Lyding1}. Fig.~\ref{fig:fig1}(b) depicts the 
relative STM height along the top of the SWNT, as indicated by the blue line in Fig. 1(a). When the SWNT is on the H-passivated
substrate\cite{a2} the height fluctuations are of the order of 0.2 \AA, and they are directly related to the underlying
honeycomb lattice of the SWNT. The dip in the apparent SWNT height stressed by the horizontal 
red line, beyond the 0.2 \AA{} fluctuations, correlates with the 
location where the SWNT traverses
the depassivated stripe (brighter region) in Fig.~\ref{fig:fig1}(a). Similar trends were reported before\cite{peternanotechnology}. Given that the substrate is 
degenerately $n$-doped, in the absence of a mechanical deformation,
 one would anticipate the negative charging of the Si surface states within
the depassivated region\cite{Dujardin} and a protrusion, rather than a dip, in the height profile: 
 to maintain constant current, the tip should retract due to the higher density of states of the Si dangling bonds. Hence the 
data in Fig.~\ref{fig:fig1}(b) provides evidence for a slight conformal deformation (of the order of 0.5 \AA) of the SWNT 
along the depassivated region. In a subsequent STM scan, the absolute 
current vs. bias was recorded ([$-2,+2$] $V$, $\Delta V$ = 20 mV) along the blue line in 
Fig.~\ref{fig:fig1}(c). In Fig.~\ref{fig:fig1}(c), we notice the apparent widening of the 
s-SWNT as the STM tip moves away from the depassivated 
region (the region within the two dashed vertical lines). This was previously observed\cite{a18} and is consistent 
with the fact that the SWNT on Si(100)-(2$\times$1):H only weakly interacts with the 
substrate\cite{a8,a13}. The d$I$/d$V$ characteristics for the SWNT shown in Fig.~1(d) are consistent with those of a s-SWNT, 
as determined in Ref.~\onlinecite{a8}. In Fig.~1(e), the absolute current vs bias, as a function of position along the 
SWNT, is shown\cite{FN2}.
 The white 
traces on Fig.~1(e) highlight the onset of the gap (upper and lower curves, at 1pA), 
as well as the midgap bias, equidistant from the conduction (upper) and valence (lower) band edges. The average onset biases are $-0.74\pm0.10$ V
and $+0.64\pm0.08$ V, when an average is made from 0 to 150 \AA; $-0.80\pm0.08$ and $0.60\pm0.06$ V, 
for an average made from 150 to 250 \AA{} in Fig.~1(e) (standard deviations are also indicated). The average 
values indicate a $\sim$0.06 V lowering of the band 
edges (and hence of the midgap) in the fully depassivated section. The presence 
of standard deviations of this order in our measurements at room 
temperature (the oscillations presented here are also seen at lower temperatures, see Ref.~\cite{Korea}) calls 
for an independent confirmation of this effect from DFT calculations.
\begin{table}
\caption{\label{tab:table2}
Parameters of the structural deformation on the SWNT (\AA).}
\begin{ruledtabular}
\begin{tabular}{lcc}
Parameter&Undoped&$n$-doped\\
\hline
$\phi z_{max}$ & 8.59 (3.0\%)    & 8.60 (3.1\%)\\
$\phi y_{max}$ & 8.63 (3.5\%)    & 8.47 (1.6\%)\\
$\phi z_{min}$ & 8.11 ($-$2.8\%) & 8.13 ($-$2.5\%)\\
$\phi y_{min}$ & 8.09 ($-$3.0\%) & 8.22 ($-$1.4\%)\\
\hline
$\Delta z_{min}$ & 2.69 & 2.75 \\
\hline
$\Delta z'_{max}$ & 1.96 & 1.94
\end{tabular}
\end{ruledtabular}
\end{table}
\begin{figure}
\includegraphics[width=.45\textwidth]{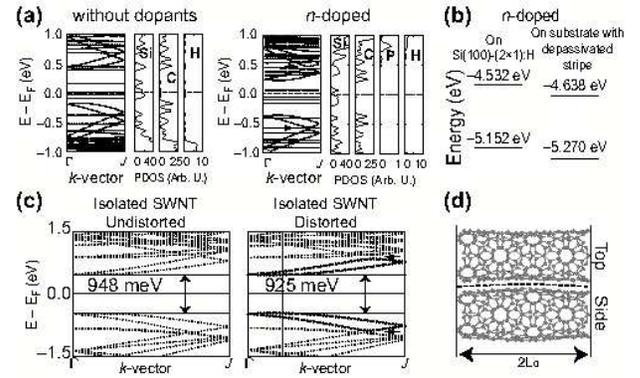}
\caption{\label{fig:Fig3}(a) Band structures and PDOS of systems shown in Fig. 2. Flat bands around the Fermi level arise from
localized states in the uppermost, unpassivated Si atoms. The highlighted PDOS of C atoms, shift downwards in
energy in the $n$-doped plot, along with the bands in the
substrate; compare to the C PDOS when no dopants are present. (b) The doping effect observed in Fig.~1(e) is independently confirmed from our calculations: Both C band edges go down by 0.11 eV
once the depassivated stripe is present in the substrate. (c) The band structure
of a 2-unit-cell long (8,4) SWNT in the absence of compression, and the
corresponding band structure of a SWNT with the parametric distortion
as in Eq.~(1). An ever slight reduction of the semiconducting gap is seen,
 as well as the breaking of the degeneracy in the points highlighted by
horizontal arrows, also present in (a) when doping is present. (d) Schematic
view of the parametric distortion (the distortion seen is larger than that in Eq.~(1) for clarity).}
\end{figure}

Calculations are performed for a (8,4) SWNT (diameter: 8.30 \AA{} and length $L_0$=11.29 \AA).  The supercell has six 
Si monolayers. The lowermost layer is passivated with H in the dihydride configuration\cite{Northrup}.
 The uppermost 
Si layer is also H passivated, but in the monohydride configuration. The area spanned by the supercell is 
43.21$\times$21.61 \AA$^2$. The supercell employed contains 1024 atoms when both Si surfaces are passivated with hydrogen, and 1012 atoms when a stripe of depassivated 
Si is formed.
The SIESTA code\cite{a19} is employed in the local density approximation (LDA) for exchange-correlation as 
parametrized by 
Perdew and Zunger\cite{PerdewZunger} from the Ceperley-Alder data\cite{a20}.
 Double-$\zeta$ plus polarization numerical atomic orbitals were used to expand the electronic wavefunctions, and a mesh cutoff of 220 Ry was employed to compute the overlap integrals.
Si dimer rows form an angle of 45$^o$ with respect to the $x$ direction, as 
seen in Fig.~2. A section 10 \AA{} wide (yellow rectangle in Fig.~2) is rendered chemically reactive by the 
removal of H atoms. Afterwards, the SWNT is placed on this substrate, crossing the depassivated section. A single 
phosphorous atom in the slab provides an $n$-type doping density of 10$^{20}$cm$^{-3}$. 
 Due to the lack of 
covalent bonding we find between s-SWNTs and Si(100), a better functional (i.e, that found in 
Ref.~\cite{vdW}) should in principle be used. We choose LDA as it well describes (by cancellation of errors)
 the spacing between the s-SWNT and Si(100) better than GGA 
functionals\cite{Nicksuggestion}. No corrections for basis set superposition error were added either. More details of the calculations can be found in Ref.~\onlinecite{a17}. Commensurability of 
the system required a non-negligible longitudinal contraction of the SWNT by 3.8\%. As seen in the inset in Fig.~2, this 
entails a reduction of 30\% in the semiconducting gap, and an increase in diameter by 0.75\% to minimize the 
additional forces caused by the longitudinal contraction\cite{FN3}.
 The entire system shown in Fig.~2 was relaxed 
employing only the $\Gamma$ point until individual forces did not exceed 0.04 eV/\AA. Because of 
periodic boundary conditions, the hydrogen stripe appears 
along the SWNT axis multiple times. This is to be contrasted with experiment, where a single 
stripe is fabricated. The SWNT displays a periodic distortion: it appears oblate with 
its maximum distortion occurring at $x_1$ and $x_2$ (Fig.~2). At $x_1$, the major axis is parallel to 
the $y-$direction; while at $x_2$ it is the minor axis that is parallel to the $y-$direction. The reason for the distortion is that the SWNT bends towards 
surface depassivated Si atoms; to relieve the most stress, this vertical elongation is accompanied by a horizontal elongation at the edges of 
the unit cell. The Si atoms in the depassivated section also protrude towards the SWNT. Specific 
values are in Table I. Although smaller than experimental values in magnitude, the DFT results are consistent with a mechanical 
distortion of the SWNT caused by the substrate (in simulations the depassivated stripe is 10 \AA{}  wide; in experiment it is about 100 
\AA{} wide). The resulting band structure, computed with a 4$\times$4$\times$1 Monkhorst-Pack $k-$point 
mesh\cite{MP}, with and without dopants is given in Fig.~3(a). Flat bands are due to dangling bonds in Si atoms 
which pin the Fermi level. The projected density of states (PDOS) of C atoms is highlighted in Fig.~3(a). For an undoped substrate, electrons from the periphery of the SWNT escape 
to the substrate, and as a result the SWNT becomes $p$-doped as the carbon HOMO level moves upward towards 
the Fermi energy (see also Ref.~\onlinecite{a17}). Upon $n$-doping of the Si substrate, the location of the carbon band edges in Fig.~3(a) moves downwards with respect to the system's Fermi level, as the Coulomb repulsion caused by 
excess electrons in the substrate suppress to some extent electron transfer from the SWNT.

 In order to provide theoretical support to the lowering of the band edges when the substrate is locally depassivated, the PDOS for the SWNT on a $n-$doped substrate with 
full H-coverage on its upper surface was obtained.
 (In this case, no further relaxation to the fully passivated H substrate was performed upon placement of the dopant atom.)
 The location of the SWNT conduction and valence band edges are shown in Fig.~3(b). It can be seen that the 
band edges shift down by 0.11 eV when the H-depassivated strip is present. The larger 
value than the one found in experiment may be due to the fact that in calculations the depassivated strip repeats infinitely along the nanotube's length. Discrepancies may also be 
due to the approximations in the calculations. (The change in the nanotube band gap of about 10 meV lays within our precision in computing the PDOS). A similar calculation was performed 
for the case when the substrate was undoped. In that case the band edges remained in their original positions even when the depassivated strip was present: The band edges in this latter case were at
$-4.69$ ($-4.69$) eV and $-5.32$ ($-5.33$) eV on the full H-passivated (partially depassivated) systems.

 The mechanical distortion in the SWNT and its relation to the electronic band structure can be understood by introducing a parametric distortion 
along the $y$ and $z$ directions to an uncompressed, isolated SWNT:
\begin{eqnarray}
y=y_0[1+0.04\cos(\pi x_0/L_0)],\nonumber\\
z=z_0[1+0.07\sin(\pi x_0/L_0)].
\end{eqnarray}
($x_0,y_0,z_0$) are the coordinates of an undistorted, uncompressed SWNT. Eq.~(1) implies a periodicity in the $x$ direction over two 
SWNT unit cells, aimed to reduce the local distortion for C atoms, while keeping a relatively small supercell. The 
dissimilar amplitude of the modulation along the $y$ and $z$ directions is responsible for the lifting of the 
degeneracy at $k-$points, highlighted by horizontal arrows in Fig.~3(a), n-doping and Fig.~3(c). 
 This parametric distortion results in a modest reduction of the semiconducting 
gap, also consistent with results from full-scale calculations (Fig.~3(a)). Fig.~3(d) schematically depicts the shape of the SWNT after a distortion as that shown in Eq.~(1) is applied. 
 The distortion also results in a shift of the nanotube's conduction and valence band edges away from the $\Gamma-$point, as is the case in Fig.~3(a).

 In conclusion, it has been shown from STM data and DFT calculations that 
partial depassivation of degenerately $n-$doped Si(100)-(2$\times$1):H produces a mechanical distortion within an adsorbed s-SWNT that slightly modifies the magnitude of
the semiconducting gap, and produces minor modifications to the band structure of the nanotube. More importantly, the partial depassivation allows for a 
slight local doping of the SWNT adsorbed on this substrate. We expect that an increase in the area of the clean stripe will result in a further 
population of electronic states at the Si(100)-SWNT interface, until the behavior described in Ref. \onlinecite{a17} is recovered.
 We thank M. Kuroda, N. A. Romero and K. Ritter for discussions. Calculations were performed on the Turing 
cluster and the Intel-64 Abe cluster at U of I. Support by the NCSA (grants TG-PHY090002 and TG-PHY090034) is acknowledged.


\begin{thebibliography}{30}
\bibitem{FN2}{The data is filtered below 1 pA, which is 
about a factor of two above the noise floor resulting from 
the tunneling preamplifier gain setting of $10^9$ V/A.}
\bibitem{FN3}{The effect of axial strain $\sigma$ on the electronic structure 
of SWNTs has been addressed before\cite{Heyd,Lyang}. We placed a (8,4) SWNT on a unit cell of dimensions 
20\AA$\times$20\AA$\times(1+\sigma)L_0$ ($\sigma$ varying from $-5$\% to $+5$\%).
 Upon 
strain, the SWNT is let to relax until forces do not exceed 0.01 eV/\AA. As a result, the SWNT diameter increases (decreases) 
 as it is longitudinally compressed (elongated).}
\end{thebibliography}
\end{document}